%% file: main.tex
\documentclass[conference]{IEEEtran}
\IEEEoverridecommandlockouts
\usepackage{cite}
\usepackage{amsmath,amssymb,amsfonts}
\usepackage{graphicx}
\usepackage{textcomp}
\usepackage{xcolor}
\usepackage{multirow}

\usepackage[utf8]{inputenc}
\usepackage{algorithmic}
\usepackage[linesnumbered,ruled,vlined]{algorithm2e}

\definecolor{darkgreen}{RGB}{0,120,0}

\usepackage[a4paper, total={184mm,239mm}]{geometry}
\def\BibTeX{{\rm B\kern-.05em{\sc i\kern-.025em b}\kern-.08em
    T\kern-.1667em\lower.7ex\hbox{E}\kern-.125emX}}

\usepackage[super]{nth}

\definecolor{myc}{cmyk}{1,0,1,0}

\begin{document}

\title{SEO: Safety-Aware Energy Optimization Framework for Multi-Sensor Neural Controllers at the Edge}

\author{\IEEEauthorblockN{Mohanad Odema, James Ferlez, Yasser Shoukry, Mohammad Abdullah Al Faruque}
\IEEEauthorblockA{Department of Electrical Engineering and Computer Science \\
University of California, Irvine, CA, USA\\}\vspace{-8truemm}}

\maketitle

\makeatletter
\renewcommand\footnoterule{%
  \kern-3\p@
  \hrule\@width.4\columnwidth
  \kern2.6\p@}
  \makeatother

\begingroup\renewcommand\thefootnote{\textsection}
\footnotetext{This work was partially supported by the NSF under awards CCF-2140154, CNS-2002405, ECCS-2139781 and by C3.ai Digital Transformation Institute.}
\endgroup

\input{sections/abstract}

\input{sections/introduction}

\input{sections/SystemModel}

\input{sections/methodology}

\input{sections/evaluation}

\input{sections/conclusion}

\bibliographystyle{IEEEtran}
\bibliography{references_short}

\end{document}

%% file: sections/abstract.tex
\begin{abstract}

Runtime energy management has become quintessential for multi-sensor autonomous systems at the edge for achieving high performance given the platform constraints. Typical for such systems, however, is to have their controllers designed with formal guarantees on safety that precede in priority such optimizations, which in turn limits their application in real settings. In this paper, we propose a novel energy optimization framework that is \textit{aware} of the autonomous system's safety state, and leverages it to \textit{regulate} the application of energy optimization methods so that the system's formal safety properties are preserved.  
In particular, through the formal characterization of a system's safety state as a dynamic processing deadline, the computing workloads of the underlying models can be adapted accordingly. For our experiments, we model two popular runtime energy optimization methods, \textit{offloading} and \textit{gating}, and simulate an autonomous driving system (ADS) use-case in the CARLA simulation environment with performance characterizations obtained from the standard Nvidia Drive PX2 ADS platform. Our results demonstrate that through a formal awareness of the perceived risks in the test case scenario, energy efficiency gains are still achieved (reaching 89.9\%) while maintaining the desired safety properties.


\end{abstract}


%% file: sections/introduction.tex
\section{Introduction} 
\label{sec:introduction}

Today, autonomous systems are capable of running high complexity neural networks (NNs) on self-sufficient edge platforms with heterogeneous hardware units (e.g., GPUs, ASICs), and integrate a wide variety of sensors (e.g., cameras, LiDAR, and IMUs) to attain a robust control performance \cite{liu2017computer}. As such, substantial computing power is required at the edge platform to enable such high performance, a requirement that goes against its other desired properties for the edge computing platform (e.g., compactness and reduced battery sizes).
Even more so, having a power-hungry computing platform can worsen the performance of other broader system functionalities, as in how an autonomous driving system (ADS) can cause reductions in a vehicle's driving range by a factor reaching 12\% \cite{lin2018architectural}.

In accordance, recent research efforts have targeted enhancing the energy efficiency of these edge platforms on both the hardware and software levels. For instance, support for processing and hardware reconfiguration has enabled effective resource management through computational workloads adjustments \cite{yi2021energy, malawade2022ecofusion}. In a similar vein, advancements in the wireless communication networking infrastructure have led to the emergence of the \textit{remote edge computing} paradigm \cite{liu2019edge, baidya2020vehicular, cui2020offloading, malawade2021sage, zamirai2020sieve}, which would equip autonomous systems with the flexibility to manage their workloads through offloading task computations to nearby servers existing at the edge of the networking infrastructure in millisecond communication latencies. 

Encouraging as it may be, the consequences of adopting such energy optimizations with regards to the safety properties of the system are quite unclear. This is a major challenge for real-world adoption scenarios as autonomous systems are required to constantly react to continuously evolving environments, prioritizing safety above all other aspects. In many cases, this is achievable in autonomous systems through \textit{provably-safe} controllers in which raw control outputs are filtered so as to be confined within the bounds of a \textit{formal} safety function, a function that is evaluated continuously through a complete, precise estimation of the corresponding system state. To give a practical example, a radar processing pipeline in an ADS can support such \textit{safety filtering} functionality, where radars inputs are processed to evaluate the safety state of the system (e.g., distance to closest obstacle), and if certain safety conditions are not satisfied (e.g., imminenet collision), the radar pipeline can override the main control pipeline to enforce safe steering or braking actions \cite{liu2017computer}. Accordingly, such a processing pipeline with critical \textit{safety} responsibilities must continuously operate at maximum performance to realize as accurate state estimates as possible for maintaining the desired control safety guarantees.


\begin{figure}[h]
\centering
{\includegraphics[,width = 0.4\textwidth]{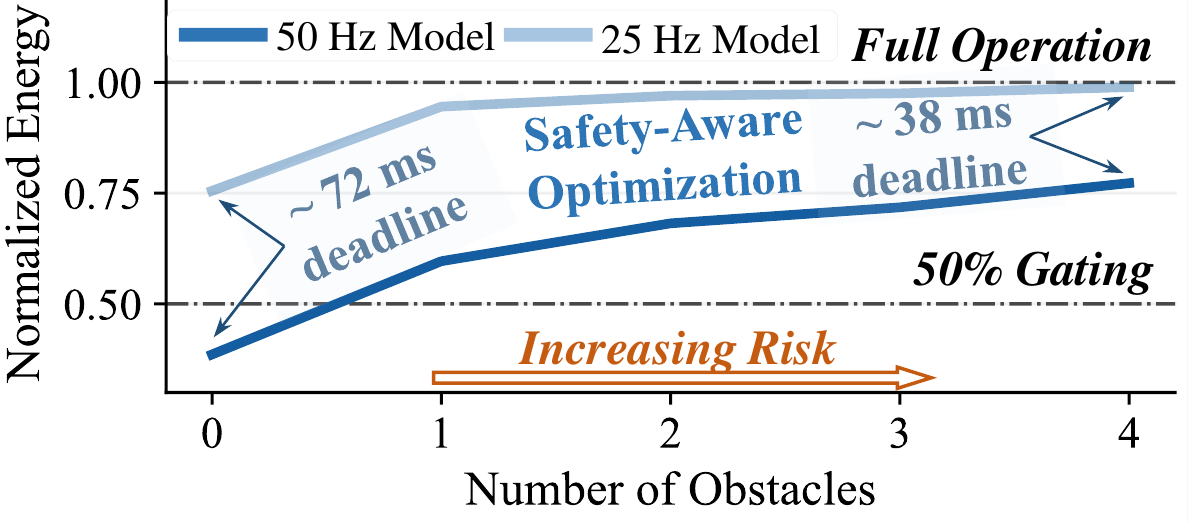}}
\vspace{-2ex}
\caption{Safety-aware gating optimization for two detector models across test runs with different number of obstacles simulated in Carla \cite{dosovitskiy2017carla}.}
\label{fig:Motivation}
\vspace{-2.5ex}
\end{figure}

\subsection{Motivational Example}
In fact, using the \textit{precise} state estimates provided through the critical safety-preserving pipelines (e.g., the Radar pipeline) and the corresponding evaluation of the safety function, we can further \textit{regulate} the application of energy optimizations onto the remaining subset of less-critical processing models in a safety-aware fashion. We showcase this in Figure \ref{fig:Motivation} through a premature example from our experiments that illustrates how this can be achieved in a \textit{formal} manner, where the test case scenario -- implemented in Carla \cite{dosovitskiy2017carla} -- involves a simulated autonomous vehicle with a pair of object detector models that support \textit{gating} of their processing models at specific time intervals for energy optimization. The detectors operate on different processing frequencies (e.g., 50 Hz and 25 Hz) to reflect heterogeneous sensors of diverse specifications and sampling frequencies \cite{gog2021pylot}. In the Figure, the horizontal axis reflects the \textit{risk} in the simulated driving scenario, represented by the number of obstacles along the vehicle's route, whereas the vertical axis represents the normalized energy consumption of the ADS under gating optimizations. As shown, the key idea is that gating optimizations are tuned based on the perceived risk on the road, i.e., safety state, through a \textit{formally-derived} safe dynamic deadline, which evaluates to \textit{lower} values at higher perceived risks (i.e., increasing number of obstacles) to prioritize robust processing over energy gains.

\subsection{Novel Contributions}
From here, we can summarize our novel contributions: 



\begin{itemize}
    \item We present SEO, a novel safety-aware energy optimization framework for multi-sensor autonomous controllers at the edge designed with specific safety properties
    \item Given the formal safety properties of an autonomous system, SEO proposes to divide the set of sensory processing models within the system into two subsets: a \textit{critical} subset that contributes to the preservation of safety guarantees, and a \textit{normal} subset
    leveraging energy optimizations. 
    \item SEO regulates the application of energy optimizations to the models in the \textit{normal} subset through a safety dynamic deadline that is estimated based on formal evaluations on the outputs from the \textit{critical} subset.
    \item We characterize the performance of the \textit{normal} processing models given dynamic safety deadlines for two popular energy optimization methods: \textit{task offloading} and \textit{gating} 
    \item Our experiments for an autonomous driving use case simulated through Carla \cite{dosovitskiy2017carla} across a variety of sensors and risk scenarios show that energy efficiency gains up to 89.9\% can be achieved under formal guarantees on safety.

\end{itemize}

\section{Related Works}

\textbf{Energy Optimizations.} Numerous methods have been proposed to effectively manage energy consumption of edge autonomous systems at runtime, most notably through: (\emph{i}) \textit{Gating} \cite{malawade2022ecofusion, lee2020accuracy} in which components of the NN pipelines, if not all, can be scaled/gated based on the corresponding system state and runtime context. (\emph{ii}) Task offloading \cite{malawade2021sage, baidya2020vehicular, odema2022testudo} in which compute-intensive kernels can be offloaded to be processed at the nearby edge computing infrastructure, enabling an effective management of the local compute resources. To date, the matter of how adopting such optimizations can affect the \textit{formal} safety properties of the autonomous system is highly understudied, especially considering the modular multi-sensory pipeline structure of today's autonomous system platforms.

\textbf{Formal Methods for NN controllers.} One research direction has been to apply formal verification techniques to asses the formal safety properties of neural network controllers \cite{sun2019formal, xiang2019reachable}. Whereas another leverages control theory concepts to augment NN controllers with formal safety guarantees on their outputs, filtering them and applying necessary corrections if needed \cite{DawsonSafeControlLearned2022, ChengEndtoEndSafeReinforcement2019}. The scope of this work aligns with the latter. Specifically, our analysis focus is on the prominent `controller-shielding' technique from that category \cite{AlshiekhSafeReinforcementLearning2017, ferlez2020shieldnn}. 

%% file: sections/SystemModel.tex
\section{System Model}

In this section, we provide the system model to formally regulate the application of energy optimizations for a controller while satisfying specific safety properties.

\subsection{Safety Guarantees for Closed-loop Controllers} \label{subsec:filter}
Let $\dot{x}=f(x, u)$ be a control system in a closed-loop with a state feedback $\pi: x \mapsto u$, where an input state, $x$, can be mapped into a control action, $u$. Let $h(x, u)$ be a real-valued function that characterizes the safety of $f$ through a binary variable, $\mathbb{S}$, based on the $x$ and $u$ estimates as follows: 
\begin{equation}
    \mathbb{S} = 
    \begin{cases}
    1,\;\;\;\;\;\;\;\;\;\;  \text{if } h(x, u) \geq 0 \\
    0,\;\;\;\;\;\;\;\;\;\;  \text{otherwise}
    \end{cases} \label{eqn:safety}
\end{equation}
where $\mathbb{S}=1$ indicates that the system is in a \textit{safe} state whenever $h$ evaluates to a \textit{non-negative} value, and an \textit{unsafe} state otherwise. In order to enforce a safe state, control outputs, $u$, are to be \textit{filtered} through a \textit{safety filter}, $\Psi$, that applies necessary corrections to $u$ if needed in order to prevent function $h(\cdot)$ from evaluating to a negative value. (i.e., $u$ remains within the bounds of the formal safety function). Thus, the \textit{filtered} control output, $u'$ can be described as:
\begin{equation}
    u' = \Psi(x, u) = 
    \begin{cases}
    u, \;\;\;\;\;\;\;\;\;\;\;\;\;\;\;\;\;\; \text{if }\mathbb{S} = 1 \\ 
    \psi(x; U), \;\;\;\;\;\;\;\; \text{otherwise}
    \end{cases} \label{eqn:shield}
\end{equation}
where $\psi$ represents a function for applying corrective behavior whenever the system is deemed to enter an unsafe state. $U$ represents the set of admissible control actions that the safety filter can apply. When a solvable function is derived for $\psi$ capturing the underlying \textit{dynamics of motion} of the physical system (e.g., the physical dynamics of rotating a steering wheel when changing steering angles) and exhibits a strong sense of uniform continuity on the control outputs, then $\dot{x}$ can be characterized as a \textit{safe} control system.



\subsection{Safe Time Intervals Characterization}
With the above safety characterization, We want to determine for a system $\dot{x}$ at $\mathbb{S} = 1$ the following: \textit{Given a state ($x_t$, $u_t$) at time $t$, denoted as $x_t$ and $u_t$, what is the maximum allowable time that $\dot{x}$ can tolerate under the same applied control action, $u_t$, before $\dot{x}$ transitions to an unsafe state ($\mathbb{S} \rightarrow 0$)?}

From equation (\ref{eqn:safety}), let $\dot{x}=f(x, u)$ be a controller in a safe state $\mathbb{S}=1$ at ($x, u$). Under the application of the same control $u$ for a certain time period, the system is expected to enter an unsafe state $\mathbb{S}=0$ at ($x', u$). Formally, if we consider $\dot{x}=f(x, u)$ and $\Psi$ enforce a strong form of uniform continuity on control outputs, that is, changes from ($x_t$, $u_t$) to the immediate next state ($x_{t+\Delta}$, $u_{t+\Delta}$) are bounded by a small constant (i.e., \textit{Lipshitz constant} in control theory). Then, we can express the maximum allowable safe time interval as such: 
\begin{equation}
    \Delta_{max} = \varphi(x, x', u)
    \label{eqn:delta}
\end{equation}
where under the application of same control value $u$, the differentiation from $x$ to $x'$ through their encompassing continuous function can be characterized in time units. At this stage, we provide the following practical example for elaboration: Let $x$ and $x'$ characterize the respective \textit{position}, \textit{velocity}, and \textit{orientation} for both an autonomous vehicle and an obstacle along its path. Then, given the vehicle's applied control values, $u$, (e.g., steering angle and throttle), we can compute the time, $\Delta_{max}$, as the \textit{time-to-collision} through numerical evaluations of $\varphi$ under the assumption that the uniform continuity property holds. In truth, we also emphasize that $x'$ is not necessarily the exact state description of the obstacle per say, but rather a characterization of its safety bound coordinates (e.g., the minimum distance to a safety sphere around the obstacle).



\subsection{Safe time Intervals as Dynamic Deadlines} \label{subsec:delta_T}
Let $\Delta_{max}$ be a real-time value representing the \textit{safety expiration time} given state ($x, u$) at a time $t$. Let $\dot{x}$ be a control system whose inputs are produced through $N$ multi-sensory processing models (e.g., $N$ neural networks) constituting the model set, $\Lambda$, contributing to the down-stream control task. Define subset $\Lambda' \subset \Lambda$ as an $N'$ subset of models in the pipeline that the safety filter, $\Psi$, does not rely on for its state estimation, $x$. This means that every $\mathcal{N}_i \in \Lambda'$ does not influence the \textit{formal} control safety guarantees. Then, $\Lambda'$ can be designated as the set of models that can benefit from incorporated runtime optimization methods whose processing workloads can be adjusted in a \emph{safety-aware} manner in accordance with the $\Delta_{max}$ values formally generated through the remainder subset of models $\Lambda'' = \Lambda - \Lambda'$.

Let each model $\mathcal{N}_i \in \Lambda'$ be associated with a single sensor, where the processing period of $\mathcal{N}_i \in \Lambda'$ is synchronized to its sensor's sampling period, denoted as $p_i$. In order to unify the time scale $\forall \mathcal{N}_i \in \Lambda'$, we define a period, $\tau$, as the base time window, and discretize the sampling periods as multiples of $\tau$:
\begin{equation}
    \forall\; \mathcal{N}_i \in \Lambda', \;\; \delta_i = 
    \begin{cases}
     \frac{p_i}{\tau},\;\;\;\;\;\;\;\;\;\;\;\;\;\; \text{if } (p_i \;\%\; \tau) == 0 \\
     \lfloor \frac{p_i}{\tau} \rfloor + 1,  \;\;\;\;\;\text{otherwise}
    \end{cases} \label{eqn:disc}
\end{equation}
Similarly, $\Delta_{max}$ can be discretized to its following multiplier:
\begin{equation}
    \delta_{max} = \lfloor \frac{\Delta_{max}}{\tau} \rfloor
\end{equation}
From here, we can regulate the application of energy optimizations for every model $\mathcal{N}_i \in \Lambda'$ to obtain its safety-aware optimized model version, $\hat{\mathcal{N}}_i$:
\begin{equation}
    \hat{\mathcal{N}}_{i[0:\delta_{max}-\delta_{i}]} = 
    \begin{cases}
    \Omega_{i[0:\delta_{max} - 2\delta_i]} + \mathcal{N}_{i(\delta_{max} - \delta_{i})} \;\;\;\text{if } \delta_i < \delta_{max}  \\
     \mathcal{N}_{i[0:\delta_{max}-\delta_i]} \;\;\;\;\;\;\;\;\;\;\;\;\;\;\;\;\;\;\;\;\;\;\;\;\;\;\text{otherwise}
    \end{cases} \label{eqn:regulate}
\end{equation}
in which $\Omega$ represents the processing model under the applied energy optimization. Thus, given a sequence of discrete time intervals indexed by $[0 : \delta_{max}-\delta_i]$, $\Omega$ can be instantiated until the last period preceding $\delta_{max}-\delta_i$ as long as $\delta_i < \delta_{max}$. After that, the original $\mathcal{N}_i$ needs to be instantiated at $\delta_{max}-\delta_i$ to meet the safety deadline at $\delta_{max}$. Otherwise, if $\delta_i \geq \delta_{max}$ (i.e., no viable optimization periods under the current deadline), $\hat{\mathcal{N}_i}$ proceeds to evaluate as the original $\mathcal{N}_i$ to maximize downstream control performance in the lesser safe states

%% file: sections/methodology.tex
\section{SEO Optimization Framework}

\begin{figure}[!tbp]
\centering
{\includegraphics[,width = 0.4\textwidth]{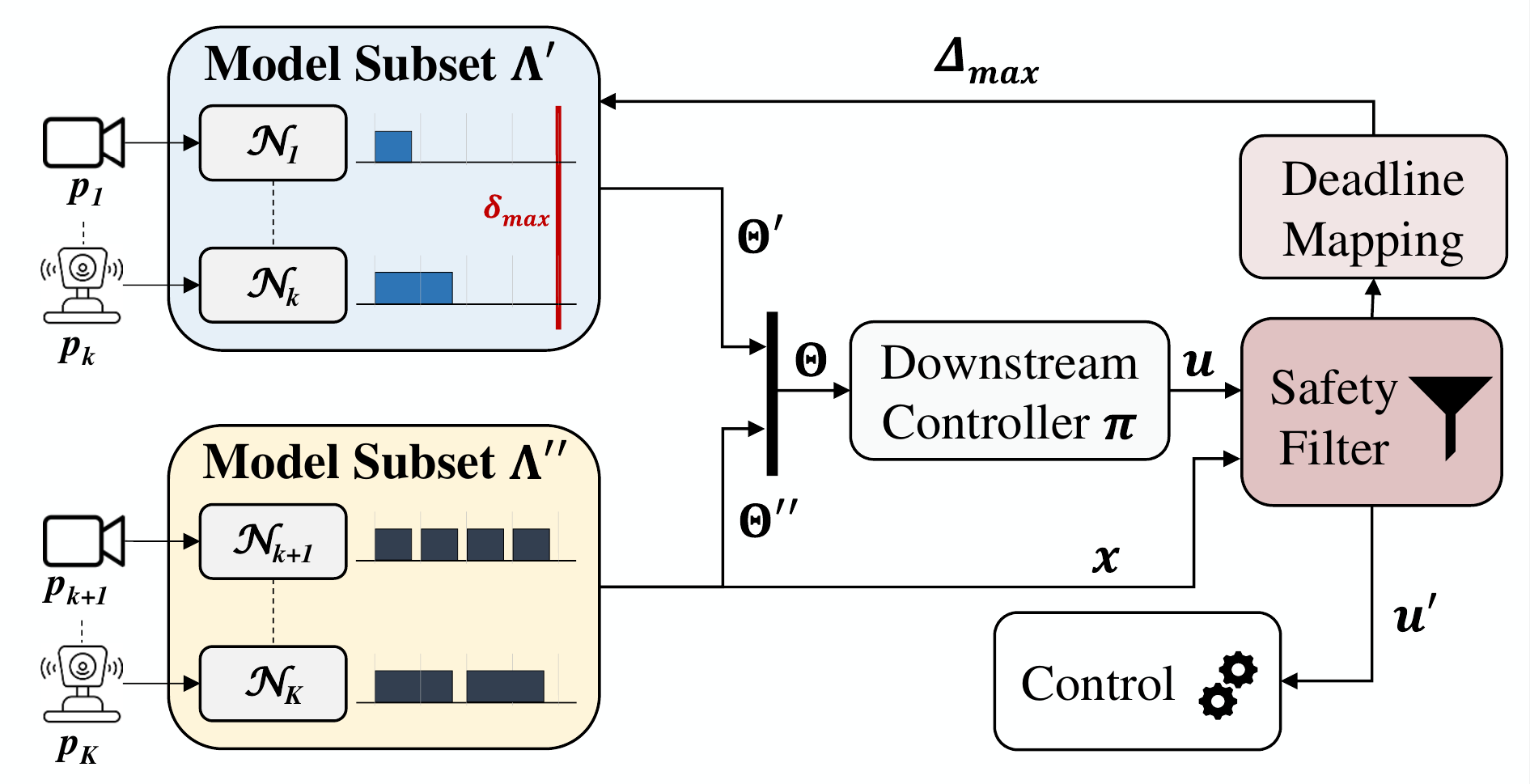}}
\vspace{-2ex}
\caption{Overview of a multi-sensor autonomous system pipeline supporting safety-aware optimizations as obtained provided through our SEO framework}
\label{fig:Method}
\vspace{-3.5ex}
\end{figure}

In this section, we present our safety-aware energy optimization framework (SEO) for an autonomous system with guarantees on safe control. Figure \ref{fig:Method} provides an illustration of how an abstract modular pipeline of a multi-sensor autonomous system would look with the supported safety-aware optimizations. 
As safety properties vary from one autonomous system to the other due to varying \textit{dynamics of motion} and \textit{control actions}, we will breakdown the different framework components below with a specific emphasis on autonomous driving systems considering how the existing literature derived and proposed methods to maintain formal safety guarantees for such systems.


\subsection{Optimization and State Estimation Subsets}


To realize energy efficiency gains while preserving the desired safety guarantees, the set of processing models deployed on an autonomous computing system are to be divided into $\Lambda'$ and $\Lambda''$ subsets according to their criticality (as defined in Section \ref{subsec:delta_T}) where critical models in $\Lambda''$ are the ones tasked with providing state estimates, $x$, to the safety controller in order to uphold the formal safety guarantees. Therefore, models in $\Lambda''$ need to be constantly operating at full processing capacity to ensure that updated state estimates, $x$, are constantly fed to the safety filter. As for $\Lambda'$, its models' evaluations are not used for safety state estimation, and thus can benefit from supported runtime performance optimizations to adapt their computational workloads. Still, these models are crucial to realize a smooth and robust end-to-end control performance along the main control pipeline, which involves the controller $\pi$ processing aggregate predictions $\Theta$ from either subset of models (see Figure \ref{fig:Method}). In other words, proposed optimizations should be applied in a context-aware, adaptive manner to limit the need for the \textit{overriding control} procedures by the safety controller. 


\subsection{The Safety Filter}\label{subsec:safety_filter}
A safety filter ensures that raw control predictions are confined within the boundaries of a safety function while accounting for the system dynamics of motion. As shown in Figure \ref{fig:Method}, the filter evaluates its safety boundaries on the corresponding state estimates generated from the model subset $\Lambda''$, and accordingly filters control predictions $u$ as $u'$ to be fed to the control unit. 
An example of such a filter is the \textit{controller shield} proposed in \cite{ferlez2020shieldnn} which was designed to filter steering angle outputs for autonomous driving control. This filter modeled the vehicle's dynamics relative to a fixed point in the plane (i.e., an obstacle), and extracted the relative distance and orientation angle as the $x$ inputs to the filter. These $x$ values are then used to evaluate the safety function $h$ with respect to the obstacle, i.e., specifying the set of safe states and control with respect to the obstacle. With that characterization, the controller shield is able to receive vehicle steering angles, and apply the necessary corrections if needed.


\subsection{Characterization of Safe Interval Times} \label{subsection: delta_T}
Given the strong sense of continuity exhibited by an autonomous system with regards to its dynamics of motion, an expression for the vehicle's progression as a function of time can be derived. Where based on the system state with respect to a reference point in the plane (e.g., an obstacle), safety expiration times, $\Delta_{max}$, can be obtained. In \cite{odema2023EnergyShield}, such a mapping function has been formally derived for the autonomous driving controller shield from the previous subsection, where the autonomous vehicle's relative states with respect to an obstacle can be mapped to a corresponding safety expiration times.
Specifically, computed $\Delta_{max}$ values based on the corresponding state (distance to obstacle and its relative orientation angle) can be leveraged as dynamic execution deadlines for the models in $\Lambda'$. For instance, a vehicle driving head on towards an obstacle within a short distance would lead to low $\Delta_{max}$ values, which in turn would cause the models in $\Lambda'$ to process inputs at near-full capacity due to the higher perceived risk.
Lastly, through enough evaluations of the safety expiration function, a low-cost proxy lookup table, denoted as \textit{\textbf{T}}$(x, u)$, is constructed to enable real-time sampling of $\Delta_{max}$ values at runtime.

\subsection{Runtime Control and Safety-Aware Optimization}

In Algorithm \ref{alg:alg}, we describe the overall runtime control loop experienced by the autonomous system with support for safety-aware optimizations. An additional notation is $y_i$ representing the input to the $i^{th}$ sensory model. \textbf{Line 3} shows the estimation of a new state, $x$, and features, $\Theta''$, from the $\Lambda''$ models to be fed to the safety component and the main controller, respectively. \textbf{Lines 4-6}, show the main control execution path in which generated controls $u$ are filtered through $\Psi$ to attain safe control actions. \textbf{Lines 7-11} indicate the start of a new safe optimization interval in which a new $\Delta_{max}$ value is sampled from \textbf{T} and discretized to $\delta_{max}$ based on the unified timing axis, whereas all $\Delta_{max}$ expiration flags are reset for the $\Lambda'$ models. The \textbf{Lines 13-21} presents our safety-aware model optimization for each involved pipeline $\mathcal{N}_i\in \Lambda'$ based on its discretized operational period,  $\delta_i$, following equation (\ref{eqn:regulate}). As detailed, the full model version, $\mathcal{N}_i$, will be invoked either when $p_i>\delta_{max}$ (no surplus optimization periods), or when $\delta_{max}$ expires. Otherwise, energy optimizations are applicable in that time step through $\Omega_n$. Prediction outputs are constantly added from each model to $\Theta'$ for $\pi$'s control outputs predictions in the following control loop. Lastly,  \textbf{Lines 22-23} show that once the optimization interval has expired for all deadlines, \textit{new$_\Delta$} flag is set to sample new $\Delta_{max}$ value in the next time step.

\begin{algorithm}[t]
	\footnotesize
	\DontPrintSemicolon
	\caption{Safe Runtime Control and Optimization}
	\label{alg:alg}
	\KwIn{Controller: $\pi$, Safety filter: $\Psi$, Lookup Table: \textbf{\textit{T}}, Base Period: $\tau$, Optimization Subset: $\Lambda'$, State Estimation Subset: $\Lambda''$ }
	\unboldmath{Initialize: \textit{n}=0, $\Delta_{max}$=0, $\Theta'$=\{\}}, \textit{new$_\Delta$}=\textit{True}\\
	\While{True}{
	\tcp{state estimation and safe control}
	$x$, $\Theta''$ $\leftarrow$ $\mathcal{N}_l(y_l, x, u) \forall \mathcal{N}_l \in \Lambda''$ \tcp*{state estimation}
	$\Theta \leftarrow aggregate({\Theta', \Theta''})$ \\
	$u \leftarrow \pi(\Theta)$   \tcp*{main control}
	$u' \leftarrow \Psi(x, u)$  \tcp*{safe control}
	\tcp{sample new safety deadline}
	\If{\unboldmath{new$_{\Delta}$ == True}}{
	$\Delta_{max} \leftarrow$ \textit{\textbf{T}}($x, u$) \tcp*{probe lookup table}
	$\delta_{max}$ = $\lfloor \frac{\Delta_{max}}{\tau} \rfloor$ \\
	\textit{n} = 0, \textit{new$_{\Delta}$} = \textit{False} \tcp*{new interval}  
	\textit{done}$_i$ == \textit{False} $\forall \mathcal{N}_i \in \Lambda'$ \tcp*{reset done flags}
	}
	\tcp{optimized safe processing}
	$\Theta'$=\{\}\\
	\For{$\mathcal{N}_i \in \Lambda'$}{
	\If{$\delta_i \geq \delta_{max}$ or \unboldmath{n} == ($\delta_{max}-\delta_{i}$)}{
	$\hat{\mathcal{N}}_{i(n)} = \mathcal{N}_{i(n)} $ 
	\tcp*{invoke processing} 
	$\theta_i$ $\leftarrow$ $\hat{\mathcal{N}}_{i(n)}(y_i)$\\
	$\Theta'$ $\cup$ $\{\theta_i\}$   \tcp*{update aggregates}
	\If{\unboldmath{n} == ($\delta_{max}-\delta_{i}$)}{
	\textit{done$_i$} = \textit{True}
	} 
	}
	\Else
    {$\hat{\mathcal{N}}_{i(n)} = \Omega_n$ \tcp*{invoke optimization}}  
	}
	\If{done$_i$ == True $\forall \mathcal{N}_i \in \Lambda'$}{
	\tcp{safe interval ended for all}
    \textit{new$_\Delta$}=\textit{True} 
	}
	\textit{n} = \textit{n} + 1
	}
\end{algorithm}

\section{Safe Energy Optimization Methods}
In this section, we describe two common methods for $\Omega$ and how they influence the operation of $\hat{\mathcal{N}}$ in equation (\ref{eqn:regulate}).
\subsection{Task Offloading}
\begin{figure}[!tbp]
\centering
{\includegraphics[,width = 0.49\textwidth]{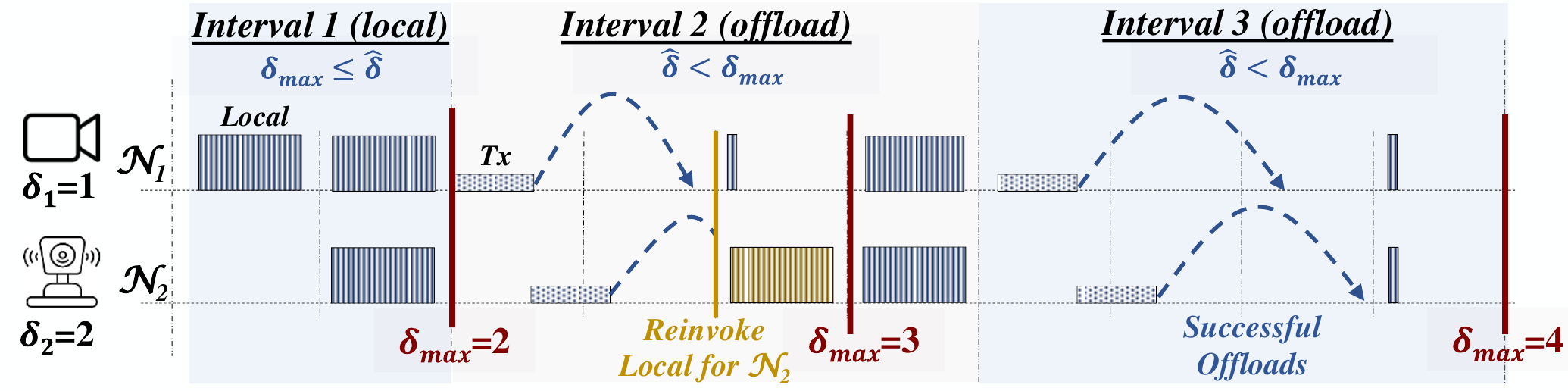}}
\vspace{-4ex}
\caption{Demonstration of task offloading under safety guarantees}
\label{fig:offload_time}
\vspace{-2ex}
\end{figure}

Through wirelessly offloading compute-intensive tasks to be processed at compute-capable servers at the edge, task offloading can offer considerable energy efficiency gains for the local computing systems \cite{baidya2020vehicular, malawade2021sage, liu2019edge}. To conduct task offloading for critical workloads (such as perception kernels affecting downstream control decisions of an autonomous vehicle), there are two aspects to be incorporated:
\begin{itemize}
    \item Server response times ($\hat{\delta}$) should be estimated to avoid offloads that are not expected to meet processing deadlines
    \item a safety fall back mechanism to re-invoke the local model if server responses after an offloading decision were delayed beyond $\hat{\delta}$ due to wireless uncertainty, and are projected to miss the critical deadline (e.g., $\delta_{max}$)
\end{itemize}
Accordingly, we demonstrate how this offloading logic can be incorporated within our primary optimization function in (\ref{eqn:regulate}). Figure \ref{fig:offload_time} provides examples of the potential experienced operational outcomes through this logic detailed below. At the start of every time interval, every model that meets the global safety deadline ($\delta_i < \delta_{max}$), proceeds to compare its $\delta_i$ against $\hat{\delta}$. If $\delta_i \leq \hat{\delta}$, then offloading is not feasible as there exists no fallback periods, and the model proceeds to evaluate locally. Otherwise, offloading is chosen with two potential outcomes: (\emph{i}) if responses are received before ($\delta_{max} - \delta_i$), then they can be applied directly as processing outputs, and thus, local compute was avoided and energy gains were realized (\emph{ii}) if ($\delta_{max} - \delta_i$) expired before receiving server responses, then the local model is instantiated to compute in the last period for safety. 

From here, given an optimizable model $\mathcal{\hat{N}}$ (see equation \ref{eqn:regulate}), we can characterize its energy consumption \textit{when offloading} (case 1 in equation \ref{eqn:regulate}) at discrete period, $n$, as follows:
\begin{equation}
    E_{\hat{\mathcal{N}}} = \underbrace{T_{tx}\cdot P_{tx}}_{E_{\Omega}} + \underbrace{\mathbb{I}[n\text{==}(\delta_{max}-\delta_{i})]\cdot T_{\mathcal{N}} \cdot P_{\mathcal{N}}} _{E_{\mathcal{N}}}
\end{equation}
where $T_{tx}$ and $P_{tx}$ are the respective transmission latency and power; $\mathbb{I}[\cdot]$ is an indicator function to invoke local processing if the guarantee on safety expires. In this case, the system incurs additional energy consumption equal to the product of $\mathcal{N}$'s local processing overheads in terms of latency, $T_{\mathcal{N}}$ and power consumption, $P_{\mathcal{N}}$. We remark that although we omitted subscript, $n$, for notational simplicity, $T_{Tx}$ and $P_{Tx}$ evaluations are dependent on it since some offloading overheads may traverse multiple windows. 

\subsection{Gating Mechanisms}
\begin{figure}[!tbp]
\centering
{\includegraphics[,width = 0.49\textwidth]{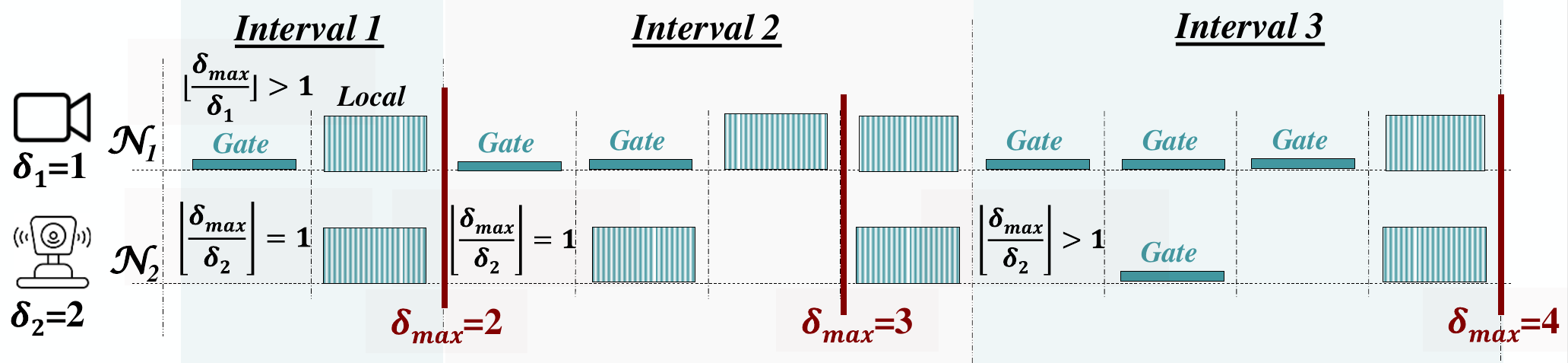}}
\vspace{-4ex}
\caption{Demonstration of gating optimizations under safety guarantees}
\label{fig:gating_time}
\vspace{-3.5ex}
\end{figure}

Gating (Figure \ref{fig:gating_time}) is another scheme for energy efficiency that benefits from the determinism offered by \textit{on-device} computing. The mechanism is straightforward in the sense given $\delta_i<\delta_{max}$, we can \textit{gate} the processing model until the final interval period for energy efficiency. Even more so, we can also gate the sensor measurements themselves when the timeline is synchronized to their sampling periods, $\tau$. In such case, we can model energy consumption for both \textit{gating} and \textit{computing} periods as:
\begin{equation}
    E_{\Omega} = \tau \cdot P_{mech},\;\;\; E_{\mathcal{N}} = \tau \cdot (P_{mech} + P_{meas}) + T_{\mathcal{N}}\cdot P_{\mathcal{N}}
    \label{eqn:sensor}
\end{equation}
in which $P_{mech}$ and $P_{measure}$ are the power drawn by the sensor due to its mechanical and measurement operations. This separation is because gating cannot be directly applied to the mechanical aspects of the sensor, such as a rotating motor, due to inertia considerations. For instance, a LiDaR sensor motor needs to keep on rotating even if sensor measurement is gated.

%% file: sections/evaluation.tex
\section{Experiments and Results}

\subsection{Experimental Setup}

We use Carla simulation environment to implement an experimental scenario similar to the one proposed in \cite{ferlez2020shieldnn} in which we have a Reinforcement Learning (RL) agent trained as an autonomous vehicle controller to travel along a 100m road that is populated with obstacles in the final third. We train the agent using the same reward function for 2000 episodes to output steering and throttle control actions. To reflect the $\Lambda''$ and $\Lambda'$ components that feed inputs into the agent, we first reuse the Variational Autoencoder in \cite{ferlez2020shieldnn} for $\Lambda''$, and deploy two pretrained ResNet-152 object detectors for $\Lambda'$, where they operate at respective periods $p=\tau$ and $p=2\tau$ to imitate sensor operational diversity \cite{gog2021pylot}. Unless otherwise stated, we set $\tau$ to 20 ms based on practical numbers from the literature and benchmark datasets \cite{gog2021pylot}.

Our forthcoming analysis for energy optimizations is conducted under both cases for when the safety component tasked with filtering steering angle outputs (recall Subsection \ref{subsec:safety_filter}) is active and inactive, referred to by respective \textit{filtered} and \textit{unfiltered}. Our main results are the average from 25 test runs in which the agent successfully completed the route without any collisions in either of the above cases. We retrieve the state estimates (i.e., distance and relative orientation) needed by the safety component directly from Carla for simplicity. 

For performance comparisons, we follow the scheme proposed in \cite{odema2022testudo} for both local and offloaded performance characterizations in terms of latency and energy consumption. Due to space considerations, we only provide a high-level overview where for the former, we deploy the ResNet-152 models on an Nvidia Drive PX2 ADS platform, and benchmark their local execution overheads using TensorRT in terms of latency and energy (17 ms latency and 7 Watts execution power consumption). For offloading, we assume a Wi-Fi link in which effective data rate values are sampled from a  Rayleigh channel distribution model with scale 20 Mbps. 

\begin{figure}[!tbp]
\centering
{\includegraphics[,width = 0.49\textwidth]{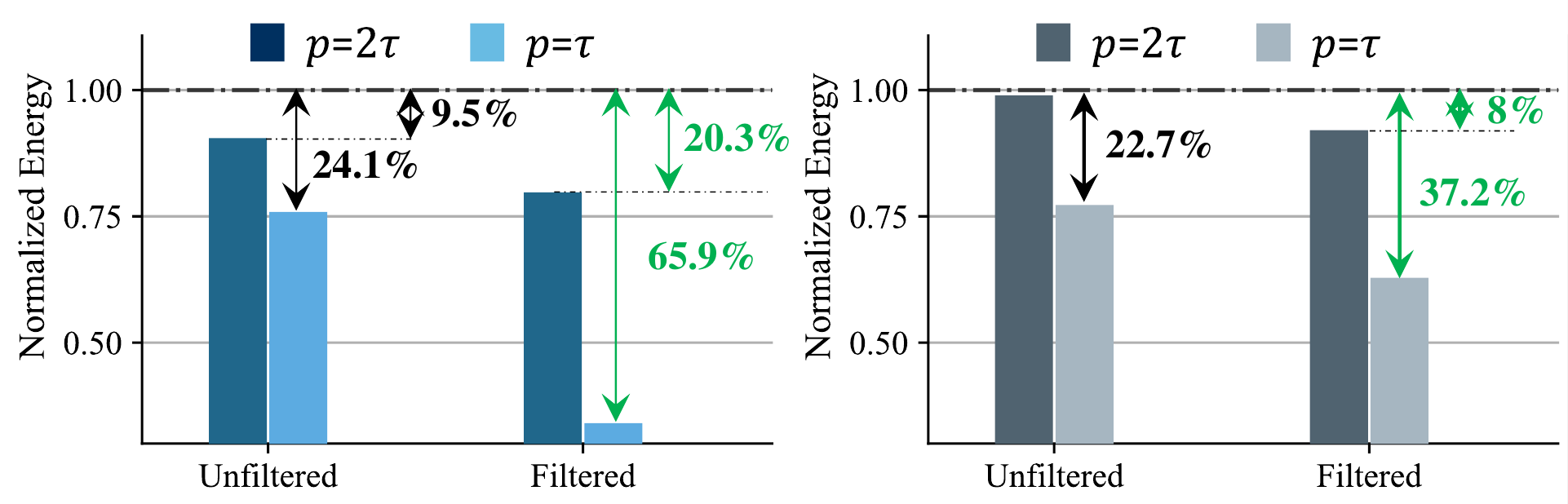}}
\vspace{-4ex}
\caption{Energy gains relative to local execution for the ResNet-152 detectors with different $p$ when \textbf{offloading (\textit{left})} and \textbf{model }\textbf{gating (\textit{right}) at $\tau=20 ms$}}
\label{fig:offload_gate_main}
\vspace{-2ex}
\end{figure}

\begin{table}[]
    \centering
    \caption{Offloading and Gating Energy Gains over local at $\tau$=25 ms}
    \vspace{-1.1ex}
    \begin{tabular}{l| c | c c c}
        Mode & Control & ($p=\tau$) gains & ($p=2\tau$) gains & Average gains \\
        \hline
        \multirow{2}{*}{Offload} & \textit{unfiltered} & 15.3\% & 7.5\% & 11.8\% \\
        & \textit{filtered} & 27.1\% & 14.1\% & 21.1\% \\
        \hline
        \multirow{2}{*}{Gating} & \textit{unfiltered} & 13.4\% & 0\% & 6.6\% \\
        & \textit{filtered} & 23.8\% & 4.3\% & 14.5\% \\
        \hline
    \end{tabular}
    \vspace{-3ex}
    \label{tab:energy_main}
\end{table}

\subsection{Energy Gains under Safety Guarantees}

To analyze the extent of energy gains under the dynamic safety execution deadlines, $\delta_{max}$, we illustrate in Figure \ref{fig:offload_gate_main} the extent of energy gains that can be realized across our two ResNet-152 detectors using \textit{offloading} and \textit{model gating} optimization methods in both the \textit{unfiltered} and \textit{filtered} cases. Based on the results, two key observations can be made: 1) models synchronized to sensors with higher sampling frequencies are naturally more likely to benefit more from proposed optimizations, as in the 65.9\% energy gains experienced by the detector at $p=\tau$ compared to the 20.3\% gains experienced by its $p=2\tau$ counterpart in the filtered offloading case, which is attributed to the former's higher prospect of optimizations under lower values of $\delta_{max}$. 2) Energy gains in the \textit{filtered} case are more than \textit{unfiltered} (e.g., 65.9\% vs 24.1\% at $p$=$\tau$ for offloading). This is mainly because the safety component forces the RL agent to maintain a healthy distance from the obstacles through effective maneuvering, which in turn causes higher values of $\delta_{max}$ being sampled and more optimizations for both models. 
We repeat our experiments in Table \ref{tab:energy_main} when varying the base period $\tau$ as a case of more limited hardware settings. As shown, considerable energy gains, are still be attainable, 21.1\% and 14.5\% on average for respective offloading and gating.



\subsection{Energy Efficiency gains under varying risk levels}

\begin{figure}[!tbp]
\centering
{\includegraphics[,width = 0.49\textwidth]{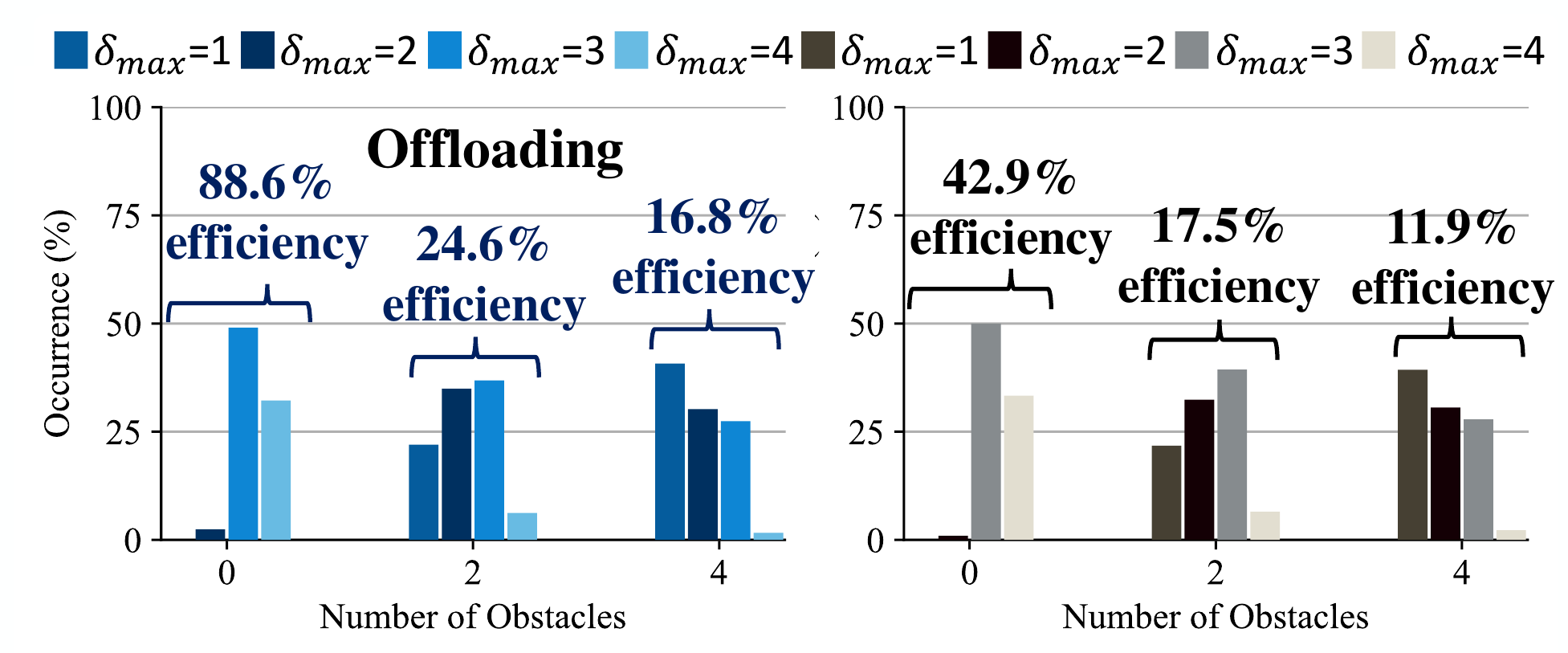}}
\vspace{-4ex}
\caption{Average $\delta_{max}$ experienced in the \textit{unfiltered} control case when varying number of obstacles for \textbf{offloading (\textit{left})} and \textbf{model gating (\textit{right})}}
\label{fig:obstacle}
\vspace{-2.5ex}
\end{figure}

To assess our approach under varying degrees of risk, we vary the number of obstacles on the vehicle's trajectory, and analyze how performance efficiency would change. Figure \ref{fig:obstacle} illustrates this for the \textit{unfiltered} case through a histogram of the sampled $\delta_{max}$ values for each variation of number of obstacles, coupled with the average energy efficiency gain over the two detectors. Across both potential optimization cases, the histogram shows that lesser values of $\delta_{max}$ are sampled more frequently as the number of obstacles increase. For instance, $\delta_{max}$=4 occurrence frequency decreases from 33.3\% to 6.48\% to 2.3\% in the model gating approach as the number of obstacles increase  from 0 to 2 to 4. That, of course, influences energy efficiency gains accordingly as indicated by the progressive drop in the average energy efficiency numbers. In Table \ref{tab:obstacle}, we also provide the results for the \textit{filtered} case. Interestingly, 
we find that the average energy gains and experienced $\delta_{max}$ values start to saturate when the number of obstacles $\geq$2. This is again attributed to minimum safety distance imposed by the safety filter leading to more evaluations of $\delta_{max}>1$. 

\begin{table}[]
    \centering
    \caption{Average Energy gains and $\delta_{max}$ at $\tau$=20 ms under obstacle variation for two combined ($p$=$\tau$) and ($p$=$2\tau$) models}
    \begin{tabular}{c | c| c c | c }
        \hline
        Control & $\#$Obst. & Offloading Gains & Gating Gains & $\delta_{max}$\\
        \hline
        \multirow{3}{*}{\textit{unfiltered}} & {0} & 88.58\% & 42.92\% & 3.67 \\
        & 2 & 24.6\% & 17.47\% & 2.29 \\
        & 4 & \textbf{16.82\%} & \textbf{11.89\%} & \textbf{1.92} \\
        \hline
        \multirow{3}{*}{\textit{filtered}} & {0} & 89.89\% & 43.82\% & 3.7 \\
        & 2 & 39.49\% & 24.26\% & 2.61 \\
        & 4 & \textbf{43.1\%} & \textbf{22.57\%} & \textbf{2.53} \\
        \hline
    \end{tabular}
    \label{tab:obstacle}
    \vspace{-3.5ex}
\end{table}

\subsection{Sensor Gating}

In this experiment, we extend our gating model analysis to encompass a broader energy consumption model of both the neural network processing model and the sensor itself (equation \ref{eqn:sensor}). Firstly, we retrieve the measurement power specifications for industry-grade sensors commonly used in autonomous systems: ZED Stereo Camera \cite{zeddatasheet}, a Navtech CTS350-X Radar \cite{navtechdatasheet}, and a Velodyne HDL-32e LiDAR \cite{velodynedatasheet}. We also specify $P_{meas}$=2.4 W for LiDAR's rotation power consumption based on common LiDAR motors \cite{malawade2022ecofusion}. The numbers are provided in Table \ref{tab:sensor}, where we compare energy gains experienced by each sensor model, both on average during the test run and and when $\delta_{max}$ was sampled equivalent to $4\tau$. As shown, energy gains for the camera pipeline achieves the best scores (37.5\% and 8.2\% on average) compared to the other sensory pipelines, this is because the absence of any residual energy consumption due to $P_{mech}$ enhances gating efficiency considerably. Between the Radar and LiDAR, we find that the RADAR is more efficient (e.g., 34.84\% vs. 32.72\% on average at $p=\tau$) as a result of the higher $P_{meas}$ (21.6 W) rating which means that it is more susceptible to benefit from sensor gating optimizations. 

\begin{table}[]
    \centering
    \caption{Sensor Gating at $\tau$=20ms for filtered Control case}
    \begin{tabular}{l|c c| c c c | c | c}
        \hline
        Sensor & $P_{meas}$ & $P_{mech}$ & Avg. Gains & $4\tau$ Gains\\
        \hline
        ZED Camera ($p$=$\tau$) & \multirow{2}{*}{1.9 W} & \multirow{2}{*}{0} & \textbf{37.5\%} & \textbf{75\%}\\
        ZED Camera ($p$=$2\tau$) & & & \textbf{8.2\%} & \textbf{50\%} \\
        \hline
        Navtech Radar ($p$=$\tau$) & \multirow{2}{*}{21.6 W} & \multirow{2}{*}{2.4 W} & \textbf{34.84\%} & \textbf{68.93\%} \\
        Navtech Radar ($p$=$2\tau$) & & & \textbf{7.57\%} & \textbf{45.53\%} \\
        \hline
        Velod. LiDAR ($p$=$\tau$) & \multirow{2}{*}{9.6 W} & \multirow{2}{*}{2.4 W} & \textbf{32.72\%} & \textbf{64.82\%} \\
        Velod. LiDAR ($p$=$2\tau$) & & & \textbf{6.9\%} & \textbf{41.91\%} \\
        \hline

        \hline
        
    \end{tabular}
    \vspace{-3.5ex}
    \label{tab:sensor}
\end{table}

%% file: sections/conclusion.tex
\section{Conclusion}

We proposed SEO a novel safety-aware energy optimization framework for multi-sensor autonomous systems at the edge that regulates how runtime energy optimizations are applied onto the involved processing pipelines. Our experiments using two common energy optimization techniques for a simulated multi-sensor autonomous vehicle in Carla environment has shown that substantial energy gains, reaching 89.9\%, can be achieved while preserving the desired safety properties.